\documentclass[12pt]{article}
\usepackage{epsfig}
\usepackage{graphicx}
\usepackage{amsthm,amsmath,amssymb}
\usepackage{amsfonts}
\usepackage{color}
\usepackage[round]{natbib}
\newcommand{\field}[1]{\mathbb{#1}}

\newcommand{\N}{\field{N}}
\newcommand{\Z}{\field{Z}}

\newcommand{\E}{\field{E}}

\theoremstyle{Conjecture} \theoremstyle{example}
\theoremstyle{remark} \theoremstyle{lemma}
\theoremstyle{definition} \theoremstyle{corol}
\theoremstyle{proposition} \theoremstyle{condition}
\newtheorem{theorem}{Theorem}

\newtheorem{lemma}{Lemma}

\def\be{\begin{equation}}
\def\ee{\end{equation}}
\def\ber{\begin{eqnarray}}
\def\eer{\end{eqnarray}}
\def\berr{\begin{eqnarray*}}
\def\eerr{\end{eqnarray*}}

\begin{document}
\centerline{\bf  Uniform  Convergence of Multivariate Spectral Density  Estimates}

\centerline{ Wei Biao Wu  }
\centerline{\it University of Chicago  }

\centerline{ Paolo Zaffaroni}
\centerline{\it Imperial College London  and University of Rome La Sapienza} 

\font\n=cmcsc10

\centerline{\today}

\begin{abstract}
We consider uniform moment convergence of lag-window spectral density estimates for univariate and multivariate stationary processes. Optimal rates of convergence are obtained under mild and easily verifiable conditions. Our theory complements earlier results which primarily concern weak or in-probability convergence. 
\end{abstract}

\section{Introduction}

Consider the $n$-dimensional stochastic process:
\be
{\bf Z }_t = ( Z_{1t},\ldots ,Z_{it},\ldots ,Z_{nt} )'
 =  {\bf R}( \ldots , {\bf \epsilon }_{t-1} , {\bf \epsilon }_{t} ), \,\,\,\,\,\,
\label{repr}
\ee
where the $ b \times 1 $ vectors $ {\bf \epsilon }_t $ are iid and ${\bf R }(.) $ is a measurable function such that ${\bf Z }_t$ exists (see Tong (1990)).  Under the above conditions $ Z_t$  is strictly stationary and ergodic  although existence of moments is not warranted. Note that  we need not  impose $n \ge b$. In fact,  we are interested in nonparametric estimation, and thus issues of invertibility and related conditions are irrelevant, unlike when considering parametric estimation methods such as maximum likelihood.
As a consequence of (\ref{repr})
\[
Z_{it} = R_i (  \ldots, {\bf \epsilon }_{t-1} , {\bf \epsilon }_{t} ), \,\,\,\, i=1,\ldots ,n,
\]
for a measurable scalar function $R_i (.)$.  In the sequel let ${\cal F}_t = (\ldots, {\bf \epsilon }_{t-1} , {\bf \epsilon }_{t} )$.

In this paper we are interested in studying  uniform convergence, in terms of distribution as well as in terms of moments, of  the  kernel  estimator  of the spectral density matrix:
\begin{eqnarray}
\label{eq:M310217}
\hat{\bf f}_T ( \lambda ) = {1 \over 2 \pi  } \sum_{u=-T+1}^{T-1} K( { u \over B_T }) e^{-  \imath  u \lambda } {\bf C }(u), \,\,\, - \pi \le \lambda < \pi ,
\end{eqnarray}
where $ \imath $ denotes the complex unit and
\[
{\bf C }(u) = {1 \over  T } \sum^{*}  {\bf Z}_t {\bf Z}_{t+u}'  \mbox{ where the  sum $ \sum^* $ is for all $t,t+u$ between $1$ and $T$},
\]
$B_T$ is the lag-window size and the kernel function satisfies
\[
K(0)=1, \mbox{ continuous and even},  \kappa = \int_{-\infty }^\infty K^2(u) du < \infty .
\]
The $(i,j)$-entry of the spectral matrix estimator is denoted by  $ \hat{f}_{Tij} ( \lambda ) $ for every $ 1 \le i,j \le n $.
Here $ \hat{\bf f}_T ( \lambda ) $ is an estimator of the true spectral density matrix which, when exists, has the form
\[
{\bf f } ( \lambda ) = {1 \over 2 \pi  } \sum_{u= - \infty }^{\infty }  e^{- \imath u \lambda } {\bf \Gamma  }(u), \,\,\, - \pi \le \lambda < \pi ,
\]
where $ {\bf \Gamma }(u) = \E( {\bf Z }_0 {\bf Z }_u' ), \, u \in \Z $ is the autocovariance matrix satisfying $ {\bf \Gamma }(-u)  = {\bf \Gamma }'(u)  $. Hereafter we assume $\E {\bf Z }_t = 0$ with, at minimum,  bounded second moment. 

To study asymptotic properties of $\hat{\bf f}_T$, we will introduce the concept of functional dependence measure. Set
\[
{\bf Z }_{t,\{0\}}  =  {\bf R } (\ldots  {\bf \epsilon }_{-1} ,  {\bf \epsilon }_{0}^* , {\bf \epsilon }_{1} \ldots   , {\bf \epsilon}_{t} ),
\]
for another iid sequence of $b \times 1 $ vector $ {\bf \epsilon }_t^* $, mutually independent from the ${\bf \epsilon}_t $.
Define $ Z_{it,\{0\}} $ accordingly. Define the $m$-dependent approximating sequence
\[
\tilde{\bf Z}_t = \E({\bf Z }_t | {\bf \epsilon }_{t-m} ,\ldots , {\bf \epsilon }_{t} ) = \E({\bf Z }_t | {\cal F }_{t-m,t} ), \, m \ge 0,
\]
with ${\cal F }_{t-m,t} = \sigma ( {\bf \epsilon }_{t-m} , \ldots, {\bf \epsilon }_t ) $ and $\tilde{Z}_{it}$ accordingly.
Set the $p$th norm, for $p>0$, equal to:
\[
\parallel {\bf Z }_t \parallel_p = (  \sum_{i=1}^n \E \mid Z_{it} \mid^p  )^{1/p} , \, \parallel {\bf Z }_t \parallel = \parallel {\bf Z }_t \parallel_2.
\]
For all $ i =1,\ldots ,n$  define the functional dependence measure
\[
\delta_{t,p}^{[i]} = \parallel Z_{it} - Z_{it,\{ 0 \}} \parallel_p ,
\]
and
\[
\Theta_{m,p}^{[i]} = \sum_{t=m}^\infty  \delta_{t,p}^{[i]},  \Psi_{m,p}^{[i]} = \left( \sum_{t=m}^\infty  ( \delta_{t,p}^{[i]} )^{p'} \right)^{1 \over p' },  \, p' = min(2,p),
\]
\[
d_{m,p}^{[i]} = \sum_{t=0}^\infty min ( \Psi_{m,p}^{[i]}, \delta_{t,p}^{[i]}).
\]
Finally, set
\begin{eqnarray*}
\delta_{t,p} &=& max_{1 \le i \le n } \delta_{t,p}^{[i]} , 
 \Theta_{m,p} = max_{1 \le i \le n } \Theta_{m,p}^{[i]} , \cr
 \Psi_{m,p} &=& max_{1 \le i \le n } \Psi_{m,p}^{[i]} , 
  d_{m,p} = max_{1 \le i \le n } d_{m,p}^{[i]} .
\end{eqnarray*}
Then $\delta_{t,p}$ quantifies of dependence of ${\bf Z }_t$ on ${\bf \epsilon }_{0}$. Our main results in the paper need conditions on the decay of $\delta_{t,p}$.

%

%
%

\section{Univariate case}

Throughout this section assume that $Z_t , t \in \Z $, is a scalar stochastic process, hence $n=1$. We also assume that $\min_\lambda f(\lambda) > 0$.  Let $\hat{f}_{T} (\cdot)$ be the lag-window estimate (\ref{eq:M310217}) and define
\begin{eqnarray}\label{eq:J05416p}
Q( \lambda) = T[\hat f_T ( \lambda) - \E \hat f_T ( \lambda)].
\end{eqnarray}
Under suitable conditions on $B_T$ and the process $(Z_t)$, we have the central limit theorem 
\begin{eqnarray}
{ {Q( \lambda)} \over \sqrt{T B_T}} \Rightarrow N(0, \kappa f^2( \lambda)), \mbox{ where } \kappa = \int K^2(u) d u.
\end{eqnarray}
For example, Anderson (1971) and Bentkus and Rudzkis (1982) dealt with linear processes and Gaussian processes, respectively and Rosenblatt (1984) considered strong mixing processes that satisfy 8th order cumulant summability conditions. Here we should consider the normalized maximum deviation
\begin{eqnarray}
 \max_{-\pi \le  \lambda < \pi} |Q( \lambda )|.
\end{eqnarray}

The following are needed on conditions on the kernel $K$ and the lag $B_T$.

\noindent
{\em
Assumption 1 (Condition 3 of Liu and Wu (2010)). $K$ is an even, bounded function with bounded support in $[-1, 1]$, $ \lim_{u \rightarrow 0} K(u) = K(0) = 1$, $ \kappa = \int_{-1}^1  K^2(u)du < 1 $ and $ \sum_{l \in \Z } \sup_{|s-l| <1 }| K(lw)-  K(sw)| = O(1)$ as $ w \rightarrow 0$.
}

\noindent
{\em
Assumption 2 (Condition 4 of Liu and Wu (2010)) There exist $ 0 < \underline{b} < b < 1 $ and $c_1, c_2 > 0 $ such that, for all large $T$,
$ c_1 T^{\underline{b}} <   B_T <  c_2 T^b $  holds.
}

\begin{theorem} \label{T1}  Let Assumptions 1 and 2 hold. Assume $ \E  Z_0 =0 , \parallel  Z_0 \parallel_p < \infty , p > 4 $ and
\be
\delta_{m,p} = O( \rho^m ) \mbox{ for some } 0 < \rho < 1 . \label{exp}
\ee
Let $ \nu^* $ be such that $ 1 \le \nu^* \le p/4 - \epsilon $, some $\epsilon > 0 $. Let $ \lambda_l^* = \pi l/B_T $. Then
\be
\left\|  \max_{0 \le l \le B_T } { T \over B_T }
 { | \hat{f}_{T} (\lambda_l^* ) - \E[ \hat{f}_{T} (\lambda_l^* )] |^2 \over   \kappa  f^2 ( \lambda_l^* )      } - 2 \log B_T + \log( \pi \log B_T )  \right\|_{\nu^*}  \rightarrow \left\| G \right\|_{\nu^*},   \label{theo1}
\ee
where $G$ denotes a Gumbel distributed random variable  with cdf $e^{-e^{-x/2}}$. 
\end{theorem}

\noindent
Remark. Condition (\ref{exp}) can be weakened to
\ber
&& d_{m,p} = O(m^{-\alpha_1 }), \,\,\, \alpha_1 > max \left[ 1/2 - (p-4)/(2 \delta p), 2 \delta /p \right] ,   \nonumber  \\
&& \Theta_{m,p} = O(m^{-\alpha_2 }), \,\,\, \alpha_2 > max \left[ 1 - (p-4)/(2 \delta p), 0 \right] \label{cond5i}
\eer
where $ B_T = O(T^b ) $ for some $ b < 1 $ by Assumption 2. Thus, when the assumptions of Theorem~\ref{T1} hold together with (\ref{cond5i}) and assuming $K(.)$ continuous with $\hat{K} (x ) = \int_{- \infty  }^\infty e^{- \imath \lambda  x } K( \lambda ) d \lambda $ satisfying $ \int_{- \infty  }^\infty | \hat{K} ( x ) | d x < \infty  $, then (\ref{theo1}) holds.

\bigskip

\noindent
{\it Proof.}  By Theorem 4 and 5 of Liu and Wu (2010)
 \[
\mathbb{ P } \left( \max_{0 \le l \le B_T } { T \over B_T }
 { | \hat{f}_{T} (\lambda_l^* ) - \E[ \hat{f}_{T} (\lambda_l^* )] |^2 \over   \kappa  f^2 ( \lambda_l^* )      } - 2 \log B_T + \log( \pi \log B_T ) \le x \right) \rightarrow e^{-e^{-x/2}}
\]
under the conditions above. Uniform convergence of the moments of the  maximum deviations  of the spectral density estimates follows once
uniform integrability of the $ \nu^* $th power of the maximum deviation is established.  We now need to  prove that for all $\nu$ with $1 \le \nu < p/2 $:
\begin{eqnarray}\label{eq:S9061123}
\left\| \max_{0 \le  \lambda \le \pi} | \hat{f}_{T} ( \lambda ) - \E[ \hat{f}_{T} ( \lambda )  ] |  \right\|_\nu = O((
B_T \log B_T / T )^{1/2}).
\end{eqnarray}
However, this is a special case of the (multivariate)  Lemma 10 reported below. QED

\section{Multivariate case}

Consider now the case of multidimensional ${\bf Z }_t$, with $n>1$.  We first need to derive the asymptotic distribution of the maximum deviations of the spectral density matrix estimator for $ {\bf f } ( \lambda ) $. Throughout this section  assume that there exists a $c_0>0$ such that $    {\bf f} ( \lambda ) - c_0 {\bf I}_n $ is positive definite  for all $ \lambda $.

\begin{theorem} \label{T2} (Theorem~5 of Liu and Wu (2010)) Let Assumptions 1 and 2  hold. Assume $ \E{\bf Z }_0 =0 , \parallel {\bf Z }_0 \parallel_p < \infty , p \ge 4 $ and
\be
\delta_{m,p} = O( \rho^m ) \mbox{ for some } 0 < \rho < 1 .
\ee
Let $ \lambda_l^* = \pi |l|/B_T $. Then for all $x \in \mathbb{R}$
 \[
\mathbb{ P } \left( \max_{0 \le l \le B_T } { T \over B_T }
 { | \hat{f}_{Tij} (\lambda_l^* ) - \E[ \hat{f}_{Tij} (\lambda_l^* )] |^2 \over   \kappa  f_{ii} ( \lambda_l^* ) f_{jj} (\lambda_l^* )     } - 2 \log B_T + \log( \pi \log B_T ) \le x \right) \rightarrow e^{-e^{-x/2}},
\]
 for every $i,j=1,..,n $.
  \end{theorem}

  \noindent
 {\it Proof.} We generalize the proof of Theorem 5 of  Liu and Wu (2010). This requires to extent a number of preliminary lemmas, presented in the Appendix. The proof then easily follows. QED

\noindent
Remark.  Theorem~\ref{T2} holds also under the weaker condition  (\ref{cond5i}).

\noindent
Remark. Theorem~\ref{T2}  permits to evaluate simultaneous confidence intervals for any subset of elements of $ \max_{0 \le l \le B_T }  {\bf f}  (\lambda_l^* )  $ via the Bonferroni method.

\medskip

\noindent
Remark. Without additional difficulties,  Theorems~1 and 2 of Liu and Wu (2010) can be generalized as follows:

\medskip

\begin{theorem} (Theorem~1 of Liu and Wu (2010)) Let Condition 1 of Liu and Wu (2010) hold. Assume $ \E{\bf Z }_0 =0 , \parallel {\bf Z }_0 \parallel_p < \infty , p \ge 2 $  and $ \Theta_{0,p} < \infty $. Let $1/B_T + B_T/T \rightarrow 0 $. Then for every $ i,j=1,..,n$
 \[
 \sup_{ \lambda \in R } \parallel  \hat{f}_{Tij} ( \lambda ) - {f}_{ij} ( \lambda ) \parallel_{p/2} \rightarrow 0.
  \]
\end{theorem}

\begin{theorem} (Theorem~2 of Liu and Wu (2010)) Let Condition 2 of Liu and Wu (2010) hold. Assume $ {\bf Z }_0 =0 , \parallel {\bf Z }_0 \parallel_4 < \infty  $
 and $ \Theta_{0,4} < \infty $. Let $1/B_T + B_T/T \rightarrow 0 $. Then for every $i,j=1,..,n$
 \[
\sqrt{ T \over B_T  } \left( \hat{f}_{Tij} ( \lambda ) - \E[ \hat{f}_{Tij} ( \lambda )] \right) \rightarrow_d N(0, \omega (  \lambda ) \kappa  f_{ii} ( \lambda ) f_{jj} ( \lambda )  ),
  \]
  for any fixed $ 0 \le  \lambda \le \pi $,  where
  $\omega (u) = 2 $ if $u /\pi \in  \Z $ and $\omega (u) = 1$ otherwise.
   The asymptotic distribution is complex normal for $ i \neq j$.
\end{theorem}

\medskip

\noindent
Remark. 
Theorem~\ref{T2} implies
\[
\max_{0 \le l \le B_T }
  | \hat{f}_{Tij} (\lambda_l^* ) - \E[ \hat{f}_{Tij} (\lambda_l^* )] |^2  = O_p \left(  { B_T \, \log B_T  \over T } \max_{0 \le l \le B_T }  f_{ii} ( \lambda_l^* ) f_{jj} (\lambda_l^* ) \right).
  \]
\noindent
Remark. If the elements of ${\bf Z}_t$ are mutually  independent, the above results hold for $p$ replaced by $p/2$.
\[
\]
\noindent
Remark. (Remark 5 of Liu and Wu (2010)) If $K(x)-1 = O(x) $ as $x \rightarrow 0 $ and $ \sum_{k \ge 1} k \delta_{k,2} < \infty $ then $\E \hat{f}_{Tij} (\lambda  ) -  f_{ij} (\lambda ) = O(B_T^{-1}) $ and we can replace $ \E \hat{f}_{Tij} (\lambda  ) $ by $  f_{ij} (\lambda ) $ for a sufficiently smooth model spectra, in particular whenever $ T \log T =o(B_T^3)$. 
  More in general, if $ \sum_{k \ge 1} k^q \delta_{k,2} < \infty $, implying that the model spectra is $q$-differentiable, then $\E \hat{f}_{Tij} (\lambda  ) -  f_{ij} (\lambda ) = O(B_T^{-q}) $. Note that under (\ref{exp}), it trivially holds that $ \sum_{k \ge 1} k^q \delta_{k,2} < \infty $ for every $ q > 1$. In this case we can replace $ \E \hat{f}_{Tij} (\lambda  ) $ by $  f_{ij} (\lambda ) $ for a sufficiently smooth model spectra, in particular whenever $ T \log T = o(B_T^{(q+1)})$. Note, however, that $q$ will also depend on the choice of the kernel $K( .)$, see Theorem 10, Chapter V, Section 4 in Hannan (1970):
\[
\lim_{x \rightarrow 0} { 1- K(x)  \over |x|^q   } = K_q < \infty .
\]
As an example, $q= \infty $ for the truncated estimator but $q=2$ for the Bartlett estimator.
\[
\]

\medskip

\noindent
Remark. We wish to have $ B_T $  as small as possible in order to achieve a quasi parametric rate but $q$ (smoothness of the spectra)  as large as possible, such that
 \[
  T \log T =o(T^{\underline{ b } (q+1)}),
 \]
 which is satisfied if
 \[
\underline{ b } (q+1) > 1 .
  \]
We now present the multivariate generalization of Theorem~\ref{T1}.

\begin{theorem} \label{T3} Under the assumptions of Theorem~\ref{T2}, for  all $ \nu^* $ such that $ 1 \le \nu^* \le p/4 - \epsilon $, some $\epsilon > 0 $:
\be
\left\|  \max_{0 \le l \le B_T } { T \over B_T }
 { | \hat{f}_{Tij} (\lambda_l^* ) - \E[ \hat{f}_{Tij} (\lambda_l^* )] |^2 \over   \kappa  f_{ii} ( \lambda_l^* ) f_{jj} (\lambda_l^* )     } - 2 \log B_T + \log( \pi \log B_T )  \right\|_{ \nu^* }  \rightarrow \left\| G \right\|_{\nu^* },   \label{theo2}
 \ee
 for every $i,j=1,..,n$, where $G$ denotes a Gumbel distributed random variable  with cdf $e^{-e^{-x/2}}$.
\end{theorem}

\noindent
{\it Proof.} Convergence of the moments follows by convergence in distribution (Theorem~\ref{T2}) and uniform integrability of the $ \nu $-th power of  $ \max_{0 \le l \le B_T }
  | \hat{f}_{Tij} (\lambda_l^* ) - \E[ \hat{f}_{Tij} (\lambda_l^* )] |^2 $. This is implied by uniform boundedness of the $ \nu  $th moments, with $\nu^* = 2\nu - \epsilon $, which follows by Lemma~\ref{Luni}. QED


\section{Appendix}

We establish here the lemmas required to proof Theorem~\ref{T3}.

\begin{lemma} (Lemma~1 of Liu and Wu (2010)) Assume $ \parallel {\bf  Z }_t \parallel_p < \infty $  for $ p > 1 $ and $\E {\bf Z }_t = 0 $. Then Lemma 1 holds for every $Z_{it}$, $i=1,\ldots ,n$.
\end{lemma}

\noindent
{\it Proof.} Trivial since each component of ${\bf Z }_t$ satisfies the assumptions of Lemma~1 of Liu and Wu (2010). QED
%
%
\[\]
\begin{lemma}  (Proposition~1 of Liu and Wu (2010)) Assume $ \parallel {\bf Z }_t \parallel_{2p} < \infty $  for $ p \ge 2 $, $\E {\bf Z}_t = 0 $ and $ \Theta_{0,2p} < \infty $. Let
\[
A_T^{[ij]} = \sum_{1 \le l  < l' \le T} \alpha_{l-l'} Z_{il} Z_{jl'},
\tilde{A}_T^{[ij]} = \sum_{1 \le l  < l' \le T} \alpha_{l-l'} \tilde{Z}_{il} \tilde{Z}_{jl'},
\]
where the $ \alpha_t $ are complex numbers.  Then
\[
{ \parallel   A_T^{[ij]} - \E A_T^{[ij]} - ( \tilde{A}_T^{[ij]} - \E \tilde{A}_T^{[ij]} )      \parallel  \over T^{1 \over 2} D_T \Theta_{0,2p} } \le C_{2p} d_{m,2p}  \mbox{ for every }  i,j=1,..,n,
\]
setting
\[
D_T = (  \sum_{s=1}^{T-1} | \alpha_s |^2 )^{1 \over 2}.
\]
\end{lemma}

\noindent
{\it Proof.} Let $ E_{it-1} = \sum_{l=1}^{t-1} \alpha_{t-l} Z_{il}, \tilde{E}_{it-1} = \sum_{l=1}^{t-1} \alpha_{t-l} \tilde{Z}_{il} $ and
\[
A_T^{[ij]*} =  \sum_{1 \le l  < l' \le T} \alpha_{l-l'} \tilde{Z}_{il} Z_{jl'} = \sum_{t=2}^T Z_{jt} \tilde{E}_{it-1}.
\]
Then
\[
\parallel {\cal P }_l ( A_T^{[ij]} - A_T^{[ij]*} ) \parallel_p \le I_l + II_l ,
\]
setting
\berr
I_l && = \parallel  \sum_{t=2}^T Z_{jt,\{l\}} \left[(  E_{it-1} - \tilde{E}_{it-1} ) - ( E_{it-1,\{l\}} - \tilde{E}_{it-1,\{l\}}    )  \right]   \parallel_p , \\
II_l && = \sum_{t=2}^T \parallel ( Z_{jt} - Z_{jt,\{l\}} )(  E_{it-1} - \tilde{E}_{it-1}    ) \parallel_p  .
\eerr
Since $ \parallel  E_{it} - \tilde{E}_{it}  \parallel_{2p} \le C_{2p} D_T \Theta_{m+1,2p}^{[i]} $ by Lemma 1, and
 $  \parallel  \tilde{Z}_{it} - \tilde{Z}_{it,\{l\}}  \parallel_{2p} \le \delta_{t-l,2p}^{[i]} $
 with $  \sum_{t=2}^T \delta_{t-l,2p}^{[i]} \le  \Theta_{0,2p}^{[i]}  $
\[
\sum_{ l = - \infty }^T II_l^2 \le C_{2p}^2 D_T^2 ( \Theta_{m+1,2p}^{[i]} )^2 \sum_{ l = - \infty }^T \Theta_{0,2p}^{[j]} (  \sum_{l'=1}^{T-1} \delta_{l'-l,2p}^{[j]}  ) \le C_{2p}^2 D_T^2 T ( \Theta_{m+1,2p}^{[i]} )^2 ( \Theta_{0,2p}^{[j]} )^2.
\]
\[
\parallel  \sum_{t=1}^{T-1}
\left[   Z_{it} - \tilde{Z}_{it}  -  Z_{it,\{l\}} + \tilde{Z}_{it,\{l\}}    ) \sum_{s=1+t}^T \alpha_{s-t} Z_{js,\{l\}}  \right]   \parallel_p
 \le 2 \sum_{t=1}^{T-1} min ( \delta_{t-l,2p}^{[i]},
  \Psi_{m+1,2p}^{[i]}   ) C_{2p} D_T \Theta_{0,2p}^{[j]},
\]
then
\[
\sum_{ l = - \infty }^T I_l^2 \le C_{2p}^2 D_T^2 ( \Theta_{0,2p}^{[j]} )^2 \sum_{ t = - \infty }^T  \Theta_{0,2p}^{[i]}
 \sum_{s=1}^{T-1} min ( \delta_{s-t,2p}^{[i]},  \Psi_{m+1,2p}^{[i]}   )   \le C_{2p}^2 D_T^2 T
( \Theta_{0,2p}^{[j]} )^2    \Theta_{m+1,2p}^{[i]} d_{m,2p}^{[i]} .
\]
Since $  \Theta_{m+1,p}^{[i]} \le d_{m,p}^{[i]} $
\berr
&& \parallel   A_T^{[ij]} -  \E A_T^{[ij]} - ( {A}_T^{[ij]*} - \E{A}_T^{[ij]*} )      \parallel_p^2 \le \sum_{l=- \infty }^T \parallel {\cal P }_l ( A_T^{[ij]} - A_T^{[ij]*} ) \parallel_p^2  \\
  && \le
   2 C_{2p}^2 D_T^2 T ( \Theta_{0,2p}^{[j]} )^2 ( d_{m,2p}^{[i]} )^2 \le 2 C_{2p}^2 D_T^2 T \Theta_{0,2p}^2  d_{m,2p}^2 .
\eerr
The same bound applies to $  \parallel   A_T^{[ij]*} - \E A_T^{[ij]*} - ( \tilde{A}_T^{[ij]} - \E \tilde{A}_T^{[ij]} )      \parallel_p^2 $. QED
\[\]
\begin{lemma} \label{3} (Proposition~2 of Liu and Wu (2010))  Assume $\E {\bf Z}_0 = 0, \parallel {\bf Z}_0 \parallel_4 < \infty , \Theta_{0,4} < \infty $. Let $ \alpha_l = \beta_l e^{\imath l \lambda } $ for
$ \lambda \in R , \beta_l \in R , 1-T \le l \le T-1 , m \in N $. Define for every $i=1,\ldots ,n$
\[
D_l^{[i]} = A_l^{[i]} - \E(  A_l^{[i]} | {\cal F }_{l-1} ), \,\,\,  A_l^{[i]} = \sum_{t=0}^\infty \E( \tilde{Z}_{it+l} | {\cal F }_l )e^{\imath t \lambda }
\]
and
\[
M_T^{[ij]} =  \sum_{t=1}^T \bar{D}_t^{[i]}  \sum_{l=1}^{t-1} \alpha_{l-t}  D_l^{[j]}, \,\,\, i,j = 1,\ldots ,n,
\]
where $ \bar{ } $ denotes complex conjugate. Then
\[
{  \parallel  \tilde{A}_T^{[ij]}    - \E \tilde{A}_T^{[ij]}    - M_T^{[ij]} \parallel  \over m^{3 \over 2} T^{1 \over 2} \parallel  Z_{i\, 0} \parallel_4   \parallel  Z_{j\, 0} \parallel_4  } \le C V_m^{1 \over 2 } (\beta )  \mbox{ for every }i,j=1,\ldots ,n.
\]
setting
\[
V_m (\beta ) = max_{ 1-T \le l \le T-1 } \beta_l^2 + m \sum_{l'=-1}^{-T-1} | \beta_{l'} - \beta_{l'-1} |^2 .
\]
\end{lemma}

\noindent
 {\it Proof.} Note that $  A_l^{[i]} = \sum_{t=0}^m \E( \tilde{Z}_{it+l} | {\cal F }_l )e^{\imath t \lambda } $ and that $ D_l^{[i]} $ is a $m$-dependent martingale difference sequence. Then, setting $U_l^{[i]} = e^{\imath (l-t) \lambda } \E( A_l^{[i]} | {\cal F }_{l-1} ) $, by summation by parts:
\berr
&& \parallel \sum_{l=1}^{t-8m} \alpha_{l-t} ( \tilde{Z}_{il} - D_l^{[i]} ) \parallel \le C m  \parallel  Z_{i \, 0} \parallel_2^{1 \over 2}  max_l \mid \beta_l \mid +
\parallel \sum_{l=1}^{t-8m} ( \beta_{l-t} - \beta_{l-t-1}   ) U_l^{[i]}  \parallel \\
&& \le C V_m^{1 \over 2 } (\beta ) m \parallel Z_{i0} \parallel_2 .
\eerr
Likewise
\[
\parallel \sum_{l=1}^{t-8m} \alpha_{l-t} ( \tilde{Z}_{il} - \bar{D}_l^{[i]} ) \parallel  \le C V_m^{1 \over 2 } (\beta ) m \parallel Z_{i0} \parallel_2 .
\]
For $ W_{1t}^{[ij]} = \tilde{Z}_{it} \sum_{l=1}^{t-8m} \beta_{l-t} e^{\imath (l-t) \lambda } ( \tilde{Z}_{jl} - D_l^{[j]} )$ then
\[
\parallel W_{1t}^{[ij]} \parallel \le C V_m^{1 \over 2} ( \beta ) m  \parallel Z_{i0} \parallel_2 \parallel Z_{j0} \parallel_2
\]
yielding
\[
\parallel  \sum_{t=1}^T W_{1t}^{[ij]}  \parallel \le \sum_{s=1}^{4m-1} \parallel \sum_{l=0}^{(T-s)/4m}  W_{1s+4ml}^{[ij]} \parallel \le C \Delta ,
\]
setting $ \Delta = max_{1 \le i,j \le n } \Delta^{[ij]} , \, \Delta^{[ij]} = V_m^{1 \over 2} ( \beta ) m^{3 \over 2} T^{1 \over 2}  \parallel Z_{i0} \parallel_2 \parallel Z_{j0} \parallel_2 $.  Except for replacing $  \parallel Z_{i0} \parallel_2 \parallel Z_{j0} \parallel_2 $ with $ \parallel Z_{i0} \parallel_4 \parallel Z_{j0} \parallel_4 $, the same bound applies to $ \parallel  \sum_{t=1}^T (  W_{2t}^{[ij]} - \E W_{2t}^{[ij]}   )  \parallel $ and $ \parallel  \sum_{t=1}^T (  W_{3t}^{[ij]} - \E W_{3t}^{[ij]}   )  \parallel $ setting
$ W_{2t}^{[ij]} = \tilde{Z}_{it} \sum_{l=t-8m+1}^{t-1} \beta_{l-t} e^{\imath (l-t) \lambda } ( \tilde{Z}_{jl} - D_l^{[j]} ) $
and $ W_{3t}^{[ij]} = ( \tilde{Z}_{it} - \bar{D}_{t}^{[i]}  )  \sum_{l=1}^{t-1} \beta_{l-t} e^{\imath (l-t) \lambda } D_l^{[j]}  $.
QED

\begin{lemma} (Lemma~2 of Liu and Wu (2010))
 Assume $ \parallel {\bf Z}_t \parallel_p < \infty $  for $ p \ge 2  $ and $\E {\bf Z}_t = 0 $. Then Lemma 2 holds for every $Z_{it}$, $i=1,\ldots ,n$.
\end{lemma}

\noindent
{\it Proof.} Trivial since each component of $Z_t$ satisfies the assumptions of Lemma~2 of Liu and Wu (2010). QED

\begin{lemma}
  (Proposition~3 of Liu and Wu (2010))
 Let ${\bf Z}_t$ be $m$-dependent with $\E{\bf Z}_t=0 , | Z_{it} | \le M $ a.s., $m \le T $ and $M \ge 1$. Let
$ S_{r,l}^{[ij]} = \sum_{t=l+1}^{l+r} Z_{it} \sum_{s=1}^{t-1} a_{T,t-s} Z_{js} $, where $l \ge 0, l+r \le T $ and assume
$  max_{1 \le t \le T } | a_{T,t} | \le K_0 $,  $  max_{1 \le t \le T }  max_{1 \le i \le n } \E Z_{it}^4 \le K_0 $ for some $K_0 > 0 $. Then for any
$x,y \ge 1 $ and $Q>0$,
\berr
&& P( | S_{r,l}^{[ij]} - \E S_{r,l}^{[ij]} | \ge x ) \le 2 e^{-y/4} + C_1 T^3 M^2 \left(   x^{-2} y^2 m^3 (M^2 + r) \sum_{s=1}^T a_{T,s}^2 \right)^Q \\
&& +  C_1 T^4 M^2 max_{1 \le i \le n } P \left(  | Z_{it} | \ge { C_2 x \over y m^2 ( M + r^{1 \over 2} )  }  \right) \mbox{ for every } i,j=1,\ldots ,n.
\eerr
\end{lemma}

\noindent
{\it Proof.} Trivial since each component of ${\bf Z}_t$ satisfies the assumptions of Proposition~3 of Liu and Wu (2010). QED
\[\]
\begin{lemma}  (Theorem~6 of Liu and Wu (2010))
Let $ a_{T,l} = b_{T,l} e^{\imath l \lambda } $, where $ \lambda \in R , b_{T,l} \in R $ with $ b_{T,l} = b_{T,-l} $ and
\[
L_T^{[ij]} = \sum_{1 \le l , l' \le T } a_{T,l-l'} Z_{il} Z_{jl'} \mbox{ and } \sigma_T^2 = \omega ( \lambda ) \sum_{r=1}^T \sum_{t=1}^T b_{T,t-r}^2 .
\]
where $ \omega ( u ) = 2 $ if $ u/\pi \in Z $  and $ \omega ( u ) = 1 $ otherwise.
Assume $ \E {\bf Z}_t = 0 , \parallel {\bf Z}_0 \parallel_4 < \infty , \Theta_{0,4} < \infty $ and
\berr
&& \max_{0 \le t \le T }  b_{T,t}^2 = o( \zeta_{T}^2 ), \, \zeta_T^2 = \sum_{t=1}^T b_{T,t}^2 , \\
&& T \zeta_T^2 = O( \sigma_T^2 ), \\
&&  \sum_{r=1}^T \sum_{t=1}^{r-1} \left|   \sum_{l=1+r}^T a_{T,r-l} a_{T,t-l} \right|^2 = o( \sigma_T^4 ), \\
&& \sum_{r=1}^T | b_{T,r} - b_{T,r-1} |^2 =o( \zeta_T^2 ).
\eerr
Then for $ 0 \le \lambda < 2 \pi $
\[
{ ( L_T^{[ij]} - \E L_T^{[ij]} ) \over \sigma_T } \rightarrow_d N ( 0, 4 \pi^2 f_{ii} ( \lambda ) f_{jj} ( \lambda ) ).
\]
\end{lemma}

\noindent
{\it Proof.} Note that
\[
L_T^{[ij]} = A_T^{[ij]} + \bar{A}_T^{[ij]} + a_{T,0} \sum_{t=1}^T Z_{it} Z_{jt} ,
\]
where  by Lemma~1
\[
\parallel \sum_{t=1}^T Z_{it} Z_{jt} - T \gamma_{ij} (0) \parallel \le C T^{1 \over 2} ( \parallel Z_{i0} \parallel_4 \Theta_{0,4}^{[j]} +
\parallel Z_{j0} \parallel_4 \Theta_{0,4}^{[i]}  ),
\]
$\gamma_{ij} (0) $ denoting the $(ij)$ entry of $ {\bf \Gamma } (0) $. It suffices to show that for any $m$
\[
{  M_T^{[ij]} +  \bar{M}_T^{[ij]} \over \sigma_T } \rightarrow_d N( 0,  N ( 0, 4 \pi^2 \tilde{f}_{ii} ( \lambda )\tilde{f}_{jj} ( \lambda ) ),
\]
and then use Bernstein's lemma, where
\[
\tilde{f}_{ii} ( \lambda ) = { 1 \over 2 \pi  } \sum_{l=-m}^m e^{\imath l \lambda } \E( \tilde{Z}_{i0} \tilde{Z}_{il} ).
\]
Since $ \parallel \sum_{t=1}^T  \bar{D}_t^{[i]} U_t^{[j]*} \parallel \le C T^{1 \over 2} \max_{1 \le t \le T } | b_{T,t} | $, setting
$ U_t^{[i]*} = \sum_{l=(t-4m+1)  \vee 1}^{t-1} a_{T,l-t}  D_l^{[i]} $, we need to show that
\[
{ 1 \over \sigma_T  }  \sum_{t=1+4m}^T ( \bar{D}_t^{[i]} U_t^{[j]\diamond} +  D_t^{[j]} \bar{U}_t^{[i] \diamond } ) \rightarrow_d N(0, 4 \pi^2 \tilde{f}_{ii} ( \lambda )\tilde{f}_{jj} ( \lambda ) ),
\]
setting $ U_t^{[i]\diamond} = \sum_{l=1}^{t-4m} a_{T,l-t}  D_l^{[i]} $. Since  $ \sum_{t=1+4m}^T \parallel \bar{D}_t^{[i]} U_t^{[j]\diamond} \parallel_4^4 \le C T \zeta_T^4 = o( \sigma_T^4 ) $ the Lindeberg condition conditions applies and  Hall and Heyde (1980) holds if
\be
{ 1 \over \sigma_T^2  }  \sum_{t=1+4m}^T \E \left( | \bar{D}_t^{[i]} U_t^{[j]\diamond} +  D_t^{[j]} \bar{U}_t^{[i] \diamond } |^2 | {\cal F }_{t-1} \right) \rightarrow_p  4 \pi^2 \tilde{f}_{ii} ( \lambda )\tilde{f}_{jj} ( \lambda ). \label{lin}
\ee
Rewriting $ \E( \cdot | {\cal F}_{t-1}) =  \sum_{r=1}^m \left( \E( \cdot | {\cal F}_{t-r} ) - \E( \cdot | {\cal F}_{t-r-1} ) \right) + \E( \cdot | {\cal F}_{t-m-1} ) $, note that for $ -m \le r \le m-1 $,
\berr
&& \parallel  \sum_{t=1+4m}^T   \left( \E[   | \bar{D}_t^{[i]} U_t^{[j]\diamond} +  D_t^{[j]} \bar{U}_t^{[i] \diamond } |^2 | {\cal F }_{t-r}  ] -    \E[   | \bar{D}_t^{[i]} U_t^{[j]\diamond} +  D_t^{[j]} \bar{U}_t^{[i] \diamond } |^2 | {\cal F }_{t-r-1}
  ]  \right)   \parallel^2  \\
 && \le 4  \sum_{t=1+4m}^T \parallel {D}_t^{[i]}  \parallel_4^4 \parallel  U_t^{[j]\diamond}  \parallel_4^4
 \le C T  \zeta_T^4 =o( \sigma_T^4 ).
\eerr
Since the $ D_t^{[i]} $ are  ${\cal F }_{t-m,t}$-measurable whilst the $U_t^{[i]\diamond } $ are
${\cal F }_{t-4m}$-measurable, $ \E ( ({D}_t^{[i]})^2 (U_t^{[j]\diamond})^2   | {\cal F }_{t-m,t})=
(U_t^{[j]\diamond})^2 \E ( ({D}_t^{[i]})^2 )      $ and (\ref{lin}) is equivalent to
\berr
&& { 1 \over \sigma_T^2  }  \sum_{t=1+4m}^T
 \left( U_t^{[i]\diamond}  U_t^{[j]\diamond}  \E ( \bar{D}_t^{[i]} \bar{D}_t^{[j]}  )
+
 \bar{U}_t^{[i]\diamond}  \bar{U}_t^{[j]\diamond}  \E ( D_t^{[i]} D_t^{[j]}  ) \right. \\
&& \left. +   | U_t^{[j]\diamond}|^2 \E ( | D_t^{[i]}| ^2 ) +
 | U_t^{[i]\diamond}|^2 \E ( | D_t^{[j]}| ^2 )   \right) \rightarrow_p || D_t^{[i]} ||^2 || D_t^{[j]} ||^2 ,
\eerr
since $ || D_t^{[i]} ||^2 = 2 \pi \tilde{f}_{ii} ( \lambda ) $. Since
 $ \parallel    \sum_{t=1+4m}^T   ( U_t^{[i]\diamond} U_t^{[j]\diamond}  ) -\E( U_t^{[i]\diamond} U_t^{[j]\diamond} ) ) \parallel = o_p( \sigma_T^2) $ and  $    \sum_{t=1+4m}^T   | \E( U_t^{[i]\diamond} U_t^{[j]\diamond}  ) | = o( \sigma_T^2) $, the result follows  noticing that
\[
     \E( |U_t^{[i]\diamond} |^2  ) =     
     \sum_{l=1}^{t-4m} b_{T,l-t}^2   \parallel  D_t^{[i]}  \parallel^2 .
\]
QED

 Set
\[
g_T^{[ij]} ( \lambda ) = L_T^{[ij]} - \E L_T^{[ij]} - \sum_{t=1}^T ( Z_{it} Z_{jt} - \E Z_{it} Z_{jt} ),
\]
and
\[
g_{T,m}^{[ij]} ( \lambda ) = \tilde{L}_T^{[ij]} - \E\tilde{L}_T^{[ij]} - \sum_{t=1}^T ( \tilde{Z}_{it} \tilde{Z}_{jt} - \E \tilde{Z}_{it} \tilde{Z}_{jt} ),
\]
noticing that unless $i=j$ then  $ g_T^{[ij]} ( \lambda )  \neq  \bar{g}_T^{[ij]} ( \lambda ) =
  g_T^{[ij]} (- \lambda )  $.
Set
\[
\tau_T = \sqrt{T B_T } / \log B_T .
\]

\begin{lemma} (Lemma~3 and Remark 7 of Liu and Wu (2010)) Let Assumptions 1 and 2 hold and  $ \E {\bf Z}_0 =0, \parallel {\bf Z }_0 \parallel_p < \infty , p > 4 $ hold. Further assume $ \delta_{s,p} = O(\rho^s ) $ for some $ 0 < \rho < 1 $.  Then for any $ 0 < {\cal C } < 1 $, there exists $ \gamma \in (0, {\cal C}) $ such that, for $m=[T^\gamma ] $, for every $i,j=1,\ldots ,n$
\[
\max_{1 \le l \le B_T} | g_T^{[ij]} ( \lambda_l^* ) - g_{T,m}^{[ij]} ( \lambda_l^* ) | = o_p( \sqrt{ T B_T / \log B_T   } ), \,\,\,
\]
\end{lemma}

\noindent
{\it Proof.} This follows precisely Liu and Wu (2010), by setting
\[
Y_{t,m}^{[ij]} ( \lambda ) = \tilde{Z}_{it} \sum_{s=1}^{t-1} a_{T,t-s} \tilde{Z}_{js}.
\]
and $ Y_{t,s^l}^{[ij]} ( \lambda ) $ accordingly, where $ s_l = [ ^{\rho^l} ], \, 1 \le l \le r , \, r \in \N $ such that $ 0  < \rho^r < {\cal C } $. Also, we replace their definition of  $ \breve{u}_r ( \lambda ) $ with
\[
\breve{u}_r ( \lambda ) = \sum_{t \in H_r } \left( ( Y_{t,s_l}^{[ij]} ( \lambda ) - Y_{t,s_{l+1}}^{[ij]} ( \lambda )  ) +   ( Y_{t,s_l}^{[ij]} ( -\lambda ) - Y_{t,s_{l+1}}^{[ij]} ( -\lambda )  )        \right).
\]
QED


\noindent
Remark. Lemma 4,5,6 of Liu and Wu (2010) extend without any additional difficulty.

\begin{lemma} (Lemma~7  of Liu and Wu (2010)) Suppose $\E {\bf Z}_0=0, \parallel {\bf Z}_0 \parallel_4 < \infty $ and $d_{T,4} = O( (\log  T )^{-2} ) $. For every $ i,j =1,\ldots ,n$ we have

\noindent
(i)
\[
| \E[ g_T^{[ij]} (\lambda_1 ) - \E g_T^{[ij]} (\lambda_1 ) ][ \bar{g}_T^{[ij]} (\lambda_2 ) -  \E\bar{g}_T^{[ij]} (\lambda_2 ) ]| = O( T B_T / (\log  B_T )^2 )
\]
 uniformly on $ \{ ( \lambda_1 , \lambda_2 ) : 0 \le \lambda_l \le \pi - B_T^{-1} (\log  B_T)^2, l=1,2 \mbox{ and } | \lambda_1 - \lambda_2 | \ge B_T (\log  B_T)^2 \} $.

\noindent
(ii)
\[
| \E[ g_T^{[ij]} (\lambda_1 ) - \E g_T^{[ij]} (\lambda_1 ) ][ \bar{g}_T^{[ij]} (\lambda_2 )-  \E\bar{g}_T^{[ij]} (\lambda_2 ) ]| = O( \alpha_T T B_T \kappa f_{ii} ( \lambda_1 ) f_{jj} (\lambda_2 ) ),
\]
 uniformly on $ \{ ( \lambda_1 , \lambda_2 ) : B_T^{-1} ( \log B_T)^2 \le \lambda_l \le \pi - B_T^{-1} (\log  B_T)^2, l=1,2 \mbox{ and } | \lambda_1 - \lambda_2 | \ge B_T^{-1}  \} $ for $a_T $ satisfying $ \limsup_{T \rightarrow \infty } \alpha_T < 1 $.

\noindent
(iii)
\[
\left| \E | g_T^{[ij]} (\lambda ) -  \E g_T^{[ij]} (\lambda ) |^2  -  4 \pi^2 T B_T f_{ii} ( \lambda ) f_{jj} (\lambda )  \right|
  = O(  T B_T (\log  B_T )^{-2} ),
\]
 uniformly on $ \{ B_T^{-1} (\log B_T)^2 \le \lambda \le \pi - B_T^{-1} (\log  B_T)^2  \} $.
\end{lemma}

\noindent
{\it Proof.} (i) and (ii).  Since $ \parallel M_T^{[ij]} (\lambda ) - N_T^{[ij]} (\lambda ) \parallel  = O( \sqrt{nm} ) $, where
\[
N_T^{[ij]} (\lambda ) = \sum_{t=1}^T \bar{D}_{t, \lambda }^{[i]}   \sum_{l=1}^{t-1-m} \alpha_{T,l-t} D_{t, \lambda }^{[j]},
\]
and $ M_T^{[ij]} ( \lambda )  =  M_T^{[ij]}  , D_{t, \lambda }^{[j]} = \bar{D}_{t }^{[j]} $ as defined in Lemma~\ref{3},
we need to show that
\[
r_{T, \lambda_1, \lambda_2 }^* = \left|  \E ( N_T^{[ij]} (\lambda_1 ) + \bar{N}_T^{[ij]} (\lambda_1 )   )
    ( \bar{N}_T^{[ij]} (\lambda_2 ) + N_T^{[ij]} (\lambda_2 )      )  \right| = O( T B_T (\log  B_T)^{-2} ),
\]
 since
\[
 \left|  \E ( M_T^{[ij]} (\lambda_1 ) + \bar{M}_T^{[ij]} (\lambda_1 )   )
    ( \bar{M}_T^{[ij]} (\lambda_2 ) + M_T^{[ij]} (\lambda_2 )      )  \right|  \le
    r_{T, \lambda_1, \lambda_2 }^* + O( T \sqrt{mB_T} + \sqrt{Tm} B_T ).
    \]
Easy calculations yield
\berr
&& r_{T, \lambda_1, \lambda_2 }^*  = \\
&& \E ( \bar{D}_{t, \lambda_1 }^{[i]} D_{t, \lambda_2 }^{[i]} ) \E ( D_{t, \lambda_1 }^{[j]} \bar{D}_{t, \lambda_2 }^{[j]} ) \sum_{t=1}^T  \sum_{l=1}^{t-m-1} K^2( (t-l)/B_T) cos((t-l)(\lambda_1 - \lambda_2)) \\
&& + \E ( \bar{D}_{t, \lambda_1 }^{[i]} \bar{D}_{t, \lambda_2 }^{[j]} ) \E ( D_{t, \lambda_1 }^{[j]} D_{t, \lambda_2 }^{[i]} ) \sum_{t=1}^T  \sum_{l=1}^{t-m-1} K^2( (t-l)/B_T) cos((t-l)(\lambda_1 + \lambda_2)).
\eerr
 Then follows the proof of Liu and Wu (2010).

 \noindent
 (iii) From (i)
\berr
&& r_{T, \lambda , \lambda }^*  = \\
&& \E ( | D_{t, \lambda }^{[i]} |^2  ) \E ( | D_{t, \lambda_1 }^{[j]} |^2 ) \sum_{t=1}^T  \sum_{l=1}^{t-m-1} K^2( (t-l)/B_T)  \\
&& + | \E ( D_{t, \lambda }^{[i]} D_{t, \lambda }^{[j]} )|^2  \sum_{t=1}^T  \sum_{l=1}^{t-m-1} K^2( (t-l)/B_T) cos((t-l)(2 \lambda )) \\
&& = O(T B_T (\log  B_T)^{-2} ) + \parallel D_{0, \lambda }^{[i]} \parallel^2 \parallel D_{0, \lambda }^{[j]} \parallel^2 T \sum_{s=-B_T}^{B_T} K^2 (s/B_T)   \\
&& = O(T B_T (\log  B_T)^{-2} ) + 4 \pi^2 \kappa  \tilde{f}_{ii} ( \lambda ) \tilde{f}_{jj} ( \lambda ),
\eerr
 where recall that $ \E  | D_{0, \lambda }^{[i]} |^2   = 2 \pi  \tilde{f}_{ii} ( \lambda ) $.
  QED

\begin{lemma} (Lemma~8  of Liu and Wu (2010)) Set $E_T = B_T - ( \log B_T)^2 $. Under the conditions of Theorem~\ref{T2} for every $i,j=1,\ldots ,n$
\[
{\cal P } \left( \max_{(\log B_T)^2 \le r \le E_T } {|  \sum_{l=1}^{k_T}\hat{u}_l^{[ij]} ( \lambda_r^* )  |^2 \over 4 \pi^2 \kappa  T B_T f_{ii} (\lambda_r )  f_{jj} (\lambda_r ) } - 2 \log(B_T)
+ \log( \pi \log B_T ) \le x                     \right) \rightarrow e^{-e^{-x/2}},
\]
with
\[
\hat{u}_l^{[ij]} ( \lambda  ) =  u_l^{[ij]} ( \lambda ) I\{| u_l^{[ij]}( \lambda )| \le \sqrt{T B_T}/(\log  B_T)^4   \} - \E \left(  u_l^{[ij]} ( \lambda ) I\{| u_l^{[ij]}( \lambda )| \le \sqrt{T B_T}/(\log  B_T)^4   \} \right),
0.
\]
and
\[
u_l^{[ij]} ( \lambda ) = \sum_{t \in H_l} ( \bar{Y}_{t,m}^{[ij]} ( \lambda ) - \E \bar{Y}_{t,m}^{[ij]} ( \lambda) ) + \sum_{t \in H_l} ( \bar{Y}_{t,m}^{[ij]} ( -\lambda) - \E \bar{Y}_{t,m}^{[ij]} ( - \lambda) ),
\]
with
\[
H_l =[ (l-1)(p_T+q_T) +1,p_T+(l-1)q_T], \,\,\, p_T = [B_T^{1+\beta  }  ] \mbox{ some small } \beta > 0  , q_T = B_T +m ,  k_T = T/(p_T + q_T )  ,
\]
and
\[
\bar{Y}_{t,m}^{[ij]} ( \lambda) = \bar{Z}_{t,m}^{[i]} \sum_{s=1}^{t-1} a_{T,t-s} \bar{Z}_{s,m}^{[j]}
\]
\[
\bar{Z}_{s,m}^{[k]} = {Z_{s,m}^{[i]}}' - \E {Z_{s,m}^{[i]}}' , \,  
 {Z_{s,m}^{[i]}}' =  Z_{s,m}^{[i]} I\{ | Z_{s,m}^{[i]}| 
  \le (TB_T)^\alpha   \}, \, \alpha  < 1/4 .
\]
\end{lemma}

\noindent
{\it Proof.}  This follows the proof of Lemma 8 in Liu and Wu (2010). QED

%

\begin{lemma} \label{Luni} Let Assumptions 1 and 2 hold. Assume $ E Z_0 = 0, \Vert Z \Vert_p < \infty ,  p > 4 $ and
\begin{align}  \label{exp2}
\delta_{m,p}^{[i]}  \le A  \rho^m   \mbox{ for some }  0 < \rho  < 1 , \,\,\,   A > 0.
\end{align}
Then  for every $1 \le i,j \le n $ and every $ 0 \le \nu \le p/2 $, setting
\[
\theta_T \equiv (T B_T \log B_T )^\frac{1}{2},
\]
one obtains, for a constant $ C_{\nu , p , b , \rho  }$  that depends only on $ \nu , p , b , \rho $,
\[
\Vert  max_{0 \le  \lambda  \le \pi } T | \hat{f}_{Tij} ( \lambda ) - E[ \hat{f}_{Tij} ( \lambda )] | \Vert_\nu    \le  C_{\nu , p , b , \rho } \theta_T .
\]
\end{lemma}
{\it Proof.}    Set  $Q_{ij}( \lambda) =  T | \hat{f}_{Tij} ( \lambda ) - E[ \hat{f}_{Tij} ( \lambda )] |$ for simplicity. Obviously (\ref{exp2}) implies $ \Theta_{m,p}^{[i]}  \le A  m^{ -\alpha } $ for any sufficiently large  $ \alpha  > 0 $.  Therefore we can assume without loss of generality that $ \alpha $ satisfies
\begin{eqnarray} \label{eq:S9061119}
 b < \alpha  p /2
 \mbox{ and } (1-2\alpha ) b < 1-4/p.
\end{eqnarray}
In fact, set $ \alpha = \max( B_1 , B_2 ) +1 $ where $B_1 = 2 b / p , B_2 = 1 - (1 - 4/p )/(2 b ) $.
In turn,   (\ref{eq:S9061119}) implies that there exists a $\beta \in (0, 1)$ such that
\begin{eqnarray} \label{eq:S9061129}
 b < \alpha  \beta  p / 2 \mbox{ and }
  (p/4-\alpha  \beta  p /2) b < p/4-1.
\end{eqnarray}
In fact,   $ \beta $ can be  obtained as $ \beta = \max(B_1,B_2)/\alpha  +  1/2 $ where $  B_1 / \alpha = 2 b / (p \alpha ) ,  B_2 / \alpha = 1/\alpha - (1- 4/p)/(2 b \alpha ) $.
Therefore $ \alpha $ and $ \beta $ only depend on $ p , b $.

  We then follow  the arguments of Theorem 10 in Xiao and Wu (2012) where, in particular, their Lemma 9 is  replaced by our Lemma 2 (see Remark S.2 in Xiao and Wu (2012b))  and their Lemma 11 and 12 are generalized using our Lemmas 2, 5 and  Corollary 1.6 and  1.7 of Nagaev (1979). It remains to show that  their result (41) is replaced by 
\begin{eqnarray}
&& \parallel \sum_{t,s=1}^T c_{s,t} ( Z_{it} Z_{js} -  \gamma_{ij} (t-s) ) \parallel_{p/2} \le   \nonumber \\
 &&  C_{p/2} {\cal  D }_T  \Big(  \sqrt{20} C_p  \sqrt{T}  \Theta_{0,p}^{[i]}  \Theta_{0,p}^{[j]} +  2^{1-2/p} ( \Theta_{0,p}^{[i]} \Vert Z_{j0}  \Vert_p  + \Theta_{0,p}^{[j]} \Vert Z_{i0}  \Vert_p  ) \Big) \nonumber  \\
 &&  \le   C_{p/2} {\cal  D }_T  \Big(  \sqrt{20} C_p  \sqrt{T} A^2 /(1 - \rho )^2 +   2^{2-2/p} C_p  A /(1 - \rho )   \Big),  \label{41}
\end{eqnarray}
where $\gamma_{ij} (u) $ denotes the $(ij)$th entry of $ {\bf \Gamma } (u) $ and
\[
{\cal D }_T^2 \equiv max \big(  \max_{1 \le s \le T } \sum_{t=1}^T c_{s,t}^2 ,   \max_{1 \le t \le T } \sum_{s=1}^T c_{s,t}^2 \big).
\]
Inequality (\ref{41})  is a consequence of  Lemma 1 and Lemma 2, as follows.  First, notice that one can rewrite
\begin{align}
\sum_{t=1}^T c_{s,t} ( Z_{it} Z_{js} -  \gamma_{ij} (t-s) )  & = \sum_{t=2}^T \sum_{s=1}^{t-1}  c_{s,t} ( Z_{it} Z_{js} -  \gamma_{ij} (t-s) )  \label{c1} \\
& + \sum_{s=2}^T \sum_{t=1}^{s-1}  c_{s,t} ( Z_{it} Z_{js} -  \gamma_{ij} (t-s) )  
+ \sum_{t=1}^T   c_{t,t} ( Z_{it} Z_{jt} -  \gamma_{ij} (0) ) \\  \nonumber
& = A_{1T}^{[ij]} + A_{2T}^{[ij]} + \sum_{t=1}^T   c_{t,t} ( Z_{it} Z_{jt} -  \gamma_{ij} (0) ).
\end{align}
We deal with the right hand side of (\ref{c1}), namely $A_{1T}^{[ij]} $, the other two terms following along the same lines.
 For simplicity  set $ E_{jt-1} = \sum_{s=1}^{t-1}  c_{s,t}  Z_{js} $ and  $ D_T = ( max_{1 \le s \le T } \sum_{t=1}^T c_{s,t}^2 )^\frac{1}{2} $.
Then, for $ {\cal P }_l ( \cdot ) \equiv E ( \cdot | {\cal F }_l ) - E ( \cdot | {\cal F}_{l-1} )  $,
\[
\parallel {\cal P }_l  A_{1T}^{[ij]}   \parallel_p \le I_l + II_l ,
\]
setting
\begin{align*}
I_l & = \parallel  \sum_{t=2}^T Z_{it,\{l\}}  \left[(  E_{jt-1}  -  E_{jt-1,\{l\}}    )  \right]   \parallel_p , \\
II_l & = \sum_{t=2}^T \parallel ( Z_{it} - Z_{it,\{l\}} )  E_{jt-1}  \parallel_p  .
\end{align*}
Since $ \parallel  E_{jt}   \parallel_{2p} \le C_{2p} D_T \Theta_{0,2p}^{[j]} $ by Lemma 1 noticing that $2p>2$,  and
 $  \parallel  \tilde{Z}_{it} - \tilde{Z}_{it,\{l\}}  \parallel_{2p} \le \delta_{t-l,2p}^{[i]} $
 with $  \sum_{t=2}^T \delta_{t-l,2p}^{[i]} \le  \Theta_{0,2p}^{[i]}  $,
\[
\sum_{ l = - \infty }^T II_l^2 \le C_{2p}^2 D_T^2 ( \Theta_{0,2p}^{[j]} )^2 \sum_{ l = - \infty }^T \Theta_{0,2p}^{[i]} (  \sum_{l'=1}^{T-1} \delta_{l'-l,2p}^{[i]}  ) \le C_{2p}^2 D_T^2 T ( \Theta_{0,2p}^{[i]} )^2 ( \Theta_{0,2p}^{[j]} )^2.
\]
 Similarly, since
\[
\parallel  \sum_{t=1}^{T-1}
[   Z_{it}  -  Z_{it,\{l\}}   ] \sum_{s=1+t}^T c_{s,t} Z_{js,\{l\}}    \parallel_p
 \le 2 \sum_{t=1}^{T-1}   \delta_{t-l,2p}^{[i]} C_{2p} D_T \Theta_{0,2p}^{[j]},
\]
then
\[
\sum_{ l = - \infty }^T I_l^2 \le 4 C_{2p}^2 D_T^2 ( \Theta_{0,2p}^{[j]} )^2 \sum_{ t = - \infty }^T  \Theta_{0,2p}^{[i]}
 \sum_{s=1}^{T-1}  \delta_{s-t,2p}^{[i]} \le  4 C_{2p}^2 D_T^2 T
( \Theta_{0,2p}^{[j]} )^2    ( \Theta_{0,2p}^{[i]} )^2  .
\]
Finally, the result follows by using $   \parallel   A_{1T}^{[ij]}   \parallel_p^2 \le C_{p}^2 \sum_{l=- \infty }^T \parallel {\cal P }_l  A_{1T}^{[ij]}   \parallel_p^2 $.
The same bound applies to $  \parallel   A_{2T}^{[ij]}  \parallel_p^2 $  where now $D_T$ must be replaced by 
$   ( max_{1 \le t \le T } \sum_{s=1}^T c_{s,t}^2 )^\frac{1}{2} $. The third term  follows by a straight application of Lemma 1.   Hence (\ref{41}) is now established.


For any $K > 1$,   there exists constants $C_{p, K, \beta }$, $C_{K, \beta }$ and $C_p$, such that, for all $x \ge
\theta_T$, we have
\begin{eqnarray}\label{eq:S9061132}
&& Pr( |Q_{ij}( \lambda)| \ge x)
 \le \\
 &&  C_{p, K, \beta }  x^{-p/2} ( \Theta_{0,p}^{[i]} \Theta_{0,p}^{[j]} )^{p/2} ( L_{T} \log T )
 + C_{K, \beta } (x^{-p/2}  ( \Theta_{0,p}^{[i]} \Theta_{0,p}^{[j]} )^{p/2} H_{T})^{K}
 +  e^{-C_p x^2 / (T B_T  ( \Theta_{0,4}^{[i]} \Theta_{0,4}^{[j]} )^2 ) }, \nonumber 
\end{eqnarray}
setting
\begin{align*}
 & L _{T} \equiv (T B_T )^\frac{p}{4} T^{-\alpha \beta \frac{p}{2}} + T B_T^{\frac{p}{2}-1-\alpha
\beta \frac{p}{ 2}} + T, \\
& H_{T} \equiv T^{1+\sqrt \beta (\frac{p}{4}-1)} B_T^\frac{p}{4}.
\end{align*}
 Specifically, the
second and the third terms in the right hand side of
(\ref{eq:S9061132}) correspond to the last two terms in inequality
(44) in Xiao and Wu (2012) whereas the first term refers to the combination of theirs (50) and (51). Hence (\ref{eq:S9061132}) follows from the generalization of 
inequalities (43), (44), (45)  in Xiao and Wu (2012).

We shall now use the large deviation inequality
(\ref{eq:S9061132}) and conclude the proof by using   $ EX^a = (1/a) \int_0^\infty x^{a-1} Pr(X>x) dx $ which holds for any positive random variable $X$ with finite  $a$th moment.
  By Theorem 7.28
in Zygmund (2002), let $Q_{ij}^* = \max_{0 \le  \lambda \le \pi}
|Q_{ij}( \lambda)|$ and $ \lambda_l = \pi l / (2 B)$, then $Q_{ij}^* \le 2
\max_{0 \le l \le 2 B} |Q_{ij} ( \lambda_l)|$ since $Q_{ij}( \lambda)$ is a trigonometric polynomial with order $B$. Hence by
 (\ref{eq:S9061132}), for a sufficiently large constant $K >
0$,
\begin{eqnarray}
&& \int_{K \theta_T}^\infty x^{\nu-1} Pr( Q_{ij}^* \ge 2 x) d x
 \le  (1+2 B_T) \int_{K  \theta_T }^\infty
  x^{\nu-1}  \max_ \lambda  Pr( |Q_{ij} ( \lambda)| \ge x) d x  \cr  \label{eq:S907906} \\
 && \hspace{-1in} \le  C_{p,K ,\beta , \nu}  (1+2B_T) 
  (\theta_T^{\nu-p/2} ( \Theta_{0,p}^{[i]} \Theta_{0,p}^{[j]} )^{p/2} L_{T} \log T + ( ( \Theta_{0,p}^{[i]} \Theta_{0,p}^{[j]} )^{p/2}  K^{-p/2} H_{T})^{K } \theta_T^{\nu-pK/2} +
     \theta_T^\nu B_T^{-C_{p,\nu} K^2/ ( \Theta_{0,4}^{[i]} \Theta_{0,4}^{[j]} )^2  }). \nonumber
\end{eqnarray}
Elementary calculations show that, under (\ref{eq:S9061129}), the
right hand side of (\ref{eq:S907906}) is $O(\theta_T^\nu)$ if we choose a
large enough $K$. Hence we have $\| Q_{ij}^* \|_\nu = O(\theta_T)$ since $ \int_0^{K \theta_T}  x^{\nu-1} Pr( Q_{ij}^* \ge 2 x) d x \le ( K \theta_T )^\nu / \nu $.
   In particular the two inequalities  in (\ref{eq:S9061129}) allow to bound the terms associated with the first and the second component of $L_{T}$. The last term of $L_{T}$ does not require any restrictions since $p/4>1$. The term involving $H_{T}$ requires $K$ large enough such that 
\begin{equation}
 A_1 = { b  \over   (p/4 -1 )(1 - \sqrt{\beta }   ) } < K ,  \label{bound1}
\end{equation}
and the third, last, term on the right hand side of (\ref{eq:S907906})   requires $K$ large enough such that
\begin{equation}
 \big(  {  ( \Theta_{0,4}^{[i]} \Theta_{0,4}^{[j]} )^2  \over C_{p,\nu} } \Big)^{1 \over 2} < K   .  \label{bound2}
\end{equation}
Since $ \Theta_{0,p}^{[i]} = \sum_{t=0}^\infty \delta_{t,p}^{[i]} \le A /( 1 - \rho ) $ for every $ i = 1, \hdots , n $, it follows that (\ref{bound2}) is implied by 
\[
A_2 = { A^2 \over   C_{p,\nu}^{1 \over 2} ( 1 - \rho )^2 }  < K .
\]
Then set $ K = max( A_1 , A_2 ) +1 $.  This implies that  $ K$  only depends  on $ \nu , p , b , \rho $.  Since the same applies to $ \alpha $ and $ \beta $, it follows that we can construct a constant $C_{\nu , p , b , \rho }$ that satisfies our statement. QED
  \[
  \]
  \[
  \]
 Remark. Lemma 10 can be   extended to the case when  $ \delta_{m,p}^{[i]} = O(  m^{- \alpha_i } )  $, for some $ \alpha_i > 0 $, by suitable modification of (\ref{eq:S9061119}), (\ref{eq:S9061129}), (\ref{bound1}) and (\ref{bound2}).

\section{References}

\par\noindent\hangindent2.3em\hangafter 1
Anderson, T.W. (1971) The Statistical Analysis of Time Series. New York: Wiley.

\par\noindent\hangindent2.3em\hangafter 1
Bentkus, R.Y. and R.A. Rudzkis (1982) On the distribution of some statistical estimates of spectral
density. Theory of Probability and Its Applications, 27, 795--814

\par\noindent\hangindent2.3em\hangafter 1
Hall, P. and  Heyde, C.C.  (1980). Martingale Limit Theory and Its Application. New York: Academic Press

\par\noindent\hangindent2.3em\hangafter 1
Hannan , E. J. (1970) Multiple Time Series. New York: Wiley.

\par\noindent\hangindent2.3em\hangafter 1
Liu, W. and  Wu, W. B. (2010) Asymptotics of Spectral Density Estimates, Econometric Theory, 26, 1218--1245

\par\noindent\hangindent2.3em\hangafter 1
Nagaev, S. V.  (1979) Large Deviations of Sums of Independent Random Variables. Annals of  Probabability,  7, 745--789.

\par\noindent\hangindent2.3em\hangafter 1
Rosenblatt, M. (1984) Asymptotic normality, strong mixing, and spectral density estimates.
Annals of Probability, 12, 1167--1180

\par\noindent\hangindent2.3em\hangafter 1
Tong, H. (1990). Non-linear Time Series: A Dynamical System Approach. Oxford University Press.

\par\noindent\hangindent2.3em\hangafter 1
Xiao, H.  and Wu, W.B. (2012)  Covariance Matrix Estimation for Stationary Time Series. Annals of Statistics,  40, 466--493.

\par\noindent\hangindent2.3em\hangafter 1
Xiao, H.  and Wu, W.B. (2012b)  Supplement to  Covariance Matrix Estimation for Stationary Time Series.

\par\noindent\hangindent2.3em\hangafter 1
Zygmund, A. (2002). Trigonometric series. Vol. I, II, Third ed. Cambridge Mathematical
Library. Cambridge: Cambridge University Press.

\end{document}